\documentclass[dvips,twocolumn,floatfix]{revtex4}
\usepackage{amsmath}
\usepackage{graphicx}
\usepackage{tikz}
\usepackage{setspace} 
\usepackage{hyperref}
\usetikzlibrary{patterns}
\def\etal{{\em et. al. }}
\begin{document}
\title{ Microscopic potential model analysis of the radiative $(n,\gamma)$ cross sections near the $Z=50$ shell closure taking part in the main s-component}
\author{Saumi Dutta }\email{saumidutta89@gmail.com}
 \author{ G. Gangopadhyay}\email{ggphy@caluniv.ac.in}
 \author{ Abhijit Bhattacharyya}\email{abhattacharyyacu@gmail.com}
\affiliation{Department of Physics, University of Calcutta, \\92, Acharya Prafulla Chandra Road, Kolkata-700009}
\begin{abstract}
The neutron capture cross sections have been studied near the $Z=50$ closed shell for a number of nuclei those take part in heavy element nucleosynthesis, the slow and the rapid neutron capture processes and the proton capture process. An optical model potential is constructed in theoretical approach by folding the density dependent M3y purely real nucleon-nucleon interaction with the target radial matter density in relativistic mean field (RMF) approach. The standard code TALYS1.8 is used for cross-section calculation. We have presented the Maxwellian-averaged capture cross-section (MACS) values and stellar neutron capture reaction rates at astrophysically relevant thermal energies and temperatures.
\end{abstract}
\maketitle

\section{Introduction}
Stellar nucleosynthesis via the charged particle induced reactions terminates when nuclear binding energy 
reaches its peak value near $A=56$. The reactions become endothermic in nature and further production of elements depends on the rates of neutron capture on stable end-products and successive $\beta$-decays. The mechanism  is  classified into two, namely, the slow neutron capture process and the rapid neutron capture process depending  on timescales of neutron captures and $\beta$-decays. There is also a minor 
contribution from p-process which occurs via the capture of protons resulting in the production of so-called p-nuclei when the path  moves near 
the proton drip line of the nucleosynthesis chart. 

An interesting region   in the path of
 heavy element nucleosynthesis exists near the proton $(Z=50)$ shell-closure. 
In this region, certain nuclei  have several long-lived, $\beta$-decaying isomers.  Abundance of some elements has contributions 
from all three processes. The exact contributions to these processes according  to the  observed abundances requires a large network calculations the key inputs of which are the neutron capture rates.   The current models for s-process and p-process, because of their inherent discrepancies, can not fully describe the origin of certain isotopes in this region of interest. The isotopic abundances can be determined from the $\sigma$N statistics, for which cross sections for all the elements in the nucleosynthesis chain have to be known with sufficient accuracy. Some elements in this region, for example, $^{114,115}$Sn have  natural abundances too low to have sufficient enrichment in the samples. Hence, experimental measurement is extremely difficult. The existing data have to be corrected for isotopic impurities.  Hence, theoretical extrapolations are in great demand in this respect.

Certain isotopes are produced entirely via s-process. These s-only isotopes are of special interest as no other nucleosynthesis mechanism  takes part in their production. As a consequence, the abundance ratios are simply their isotopic ratios and hence they are very useful in constraining certain parameters regarding s-process study. Especially, nuclei in the concerned region are synthesized in the main component of s-process, which, in general, occurs in low-mass thermally pulsing asymptotic giant branch (AGB) stars. The neutron flux in the main component is sufficient so that the steady flow equilibrium is achieved and hence, the elements can be produced to their saturation abundances.  In contrast to the main component, uncertainty in cross section of one isotope affects only that particular isotope and not the entire abundance distribution and hence, the abundance pattern does not suffer from the so-called propagation effect. 
Branchings   in the s-process path occur whenever the neutron capture rate becomes comparable to the corresponding $\beta$-decay rate. Then a competition takes place between the two processes and one can define the branching ratio which is defined as the ratio of neutron capture rate to the sum of all decay and capture rates. 

The cross sections for long-lived fission products (LLFPs), produced in fission reactors, for example, for $^{129}$I, is required in nuclear transmutation technology in which the harmful LLFPs are reduced in amount by converting into stable or short-lived nuclei. 
 In addition, some stable isotopes are also produced in fission reactors. Their cross sections are also very important for isotopic separation of LLFP if the transmutation system is not adapted to it. 

In the high-temperature stellar environment, the neutrons are thermalized by collisions.
The experimental neutron spectra, produced, in principle, by three reactions, e.g., $^{7}$Li$(p,n)^{7}$Be, $^{18}$O$(p,n)^{18}$F, and $^{13}$C$(\alpha,n)^{16}$O at thermal energies of 25, 5, and 8 keV, respectively. However, modern network codes coupled with  stellar models require cross sections  at several higher energies up to $\sim$ 100 keV during various phases of stellar burning. In such scenarios, the cross sections at higher energies have to be extrapolated from statistical model calculations.

Thus, in general, experiments can not be performed at all astrophysical energies for all concerned isotopes. Further, it is not possible to measure the stellar reaction rates  under normal laboratory conditions. 
 Hence, all the complicated reaction mechanisms activated in interstellar environments are impossible to measure in normal laboratory conditions and theoretical extrapolations are required. On the other hand, experimental information  is necessary to test the validity and to impose the requirement of further developments of theoretical models. In fact, only a close collaboration between theory and experiment  will help in getting  the complete picture of stellar heavy element nucleosynthesis.

 The current work presents Hauser-Feshbach statistical model  calculations of reaction cross sections and rates at astrophysically important energies for various nuclei around the $Z=50$ shell closure.
 The current paper is organized as follows. First, we have briefly described the theory.  In the results section, the  $(n,\gamma)$ cross sections for a  number of elements starting from indium to xenon are compared with the available experimental results. After that, the   Maxwellian-averaged  cross sections, first at 30 keV and then at several other energies are presented along with the MOST predictions and available experimental data. We have also presented the astrophysical reaction rates for some selected important isotopes around the Sn shell closure. Finally, we have discussed our work.

\section{Theoretical framework}
The microscopic optical model potential is widely used to describe the absorption and scattering phenomena. A complex microscopic optical model potential simplifies the complicated many-body descriptions of nucleon-nucleus interaction by an average one-body mean-field potential. It divides the incident flux into a part describing  elastic scattering and another part describing other non-elastic channels. The solution of the Schr{\"o}dinger equation with this potential then yields angular distribution as well as total reaction cross sections. 


The basic building block in constructing an optical model potential is the nucleon-nucleon interaction.
 We have chosen the standard real density dependent M3Y (DDM3Y) interaction \cite{ddm3y} and folded it with the target radial matter densities, obtained from relativistic mean field model. The integration  is done in spherical coordinates \cite{n82}.
 A spin-orbit interaction term with energy dependent phenomenological potential depths  is also included according to Scheerbaum's prescription \cite{scheerbaum}. 
In our previous studies, this potential has been found to describe the elastic scattering as well as proton and neutron capture reaction phenomena at  low
 energies \cite{gg_cl1,gg_cl2,gg_cl3,55-60,110-125,40-55,n82,n50}. 
In the present work, this potential has been used to calculate the radiative 
 neutron capture  cross sections in and around the $Z=50$  shell-closure. 

The baryonic matter density is extracted in the relativistic-mean-field (RMF) approach. 
The standard FSU Gold lagrangian density  \cite{fsugold} with a definite set of parameters \cite{n82} is used to describe the RMF theory.
 Our RMF model has previously been found to reproduce the experimental binding energies and root-mean-square charge radius values in this region of interest \cite{110-125}.

  The cross sections are calculated in  compound nuclear  Hauser-Feshbach (HF) formalism using the statistical model reaction code TALYS1.8 \cite{talys2}. Various input parameters play crucial roles in statistical model calculation. 
 The level densities are
 taken from Goriely's microscopic calculations \cite{ldmodel}. In the case of  
radiative capture cross sections, the exit channel deals with one or more $\gamma$-rays. 
 The strength function which describes the transmission, depends on the $\gamma$-ray 
energy and also on the energy and width of giant dipole resonances. We have taken the dominant 
 E1 $\gamma$-ray strength function from the  Hartree-Fock-Bogolyubov calculation \cite{e1strength}.   The transmission coefficients are obtained from the microscopic optical model potential as the function of phase shifts. The complex phase shifts  are represented in terms of the logarithmic derivative of the wave functions obtained from the solution of Sch{\"o}dinger wave equation. Width fluctuation correction is also included in our calculation according to Moldauer's formula. These are mainly the correlation factors with which all partial channels for  incoming and  outgoing particles have to be multiplied. These renormalization factors redefine the transmission coefficients by redistributing the total width over all possible channels in order to conserve the total cross section. The major effect is to enhance the elastic channel and weak channel over the dominant one.
 \begin{figure*}
\includegraphics[scale=1.05]{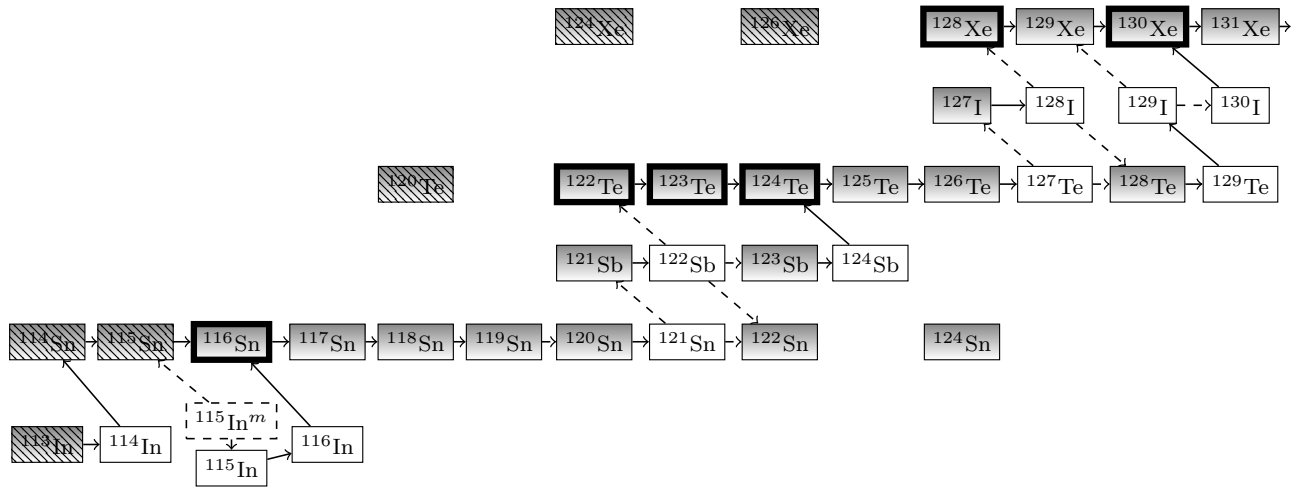}
\caption{(Color online) S-process path near Sn-Sb-Te region. The shaded rectangles represent stable  isotopes. The weak branchings are designated by dashed lines. Rectangles with thick borders denote s-only isotopes and those filled with patterns represent p-nuclei.
\label{s-path}}
\end{figure*}
 The peak and width of the distribution depends on the centrifugal quantum number $(l)$. This $l$ accounts for the contribution of different partial waves. Hence centrifugal barrier does play role in the case of neutron induced reactions by shifting the peak and width slightly. We have calculated the $(n,\gamma)$ cross sections in the energy range from 1 keV to 1 MeV. 

Neutrons are easily thermalized in a stellar high-temperature environment. Hence, Maxwellian-averaged  cross sections are obtained by averaging the total $(n,\gamma)$ cross sections over the Maxwell-Boltzmann distribution as follows.
\begin{equation}
<\sigma>=\frac{2}{\sqrt\pi} \frac{ \int_{0}^{\infty} \sigma(E_{n}) E_{n} exp(-E_{n}/kT) dE_{n}}{\int_{0}^{\infty} E_{n} exp(-E_{n}/kT) dE_{n}}
\end{equation}

Here, $E_{n}$ is the energy in center-of-mass frame, $K$ is Boltzmann constant, and $T$ is the temperature. Similarly, Maxwellian-averaged stellar reaction rates can be obtained by considering  thermodynamical equilibrium between the compound nuclear cross sections of nuclei existing in ground states as well as in different excited states.  
More details on theoretical formalism are available in Dutta \etal \cite{n82,n50}. 

Classical s-process calculation  prefers the MACS value at the energy of 30 keV. However, in recent days, more improved network models coupled with stellar hydrodynamics, those explicitly take care of branching analysis,  demand MACS values over a range of thermal energies. In most of the cases, especially the measurement by activation technique, it is not always possible to have MACSs over the entire required range and in those cases extrapolations of calculated ones are very necessary.  For this reason, we have calculated the MACSs from 5  to 100 keV for a few  selected important isotopes those do not have experimental MACSs available.  

\begin{figure}
\includegraphics[scale=0.600]{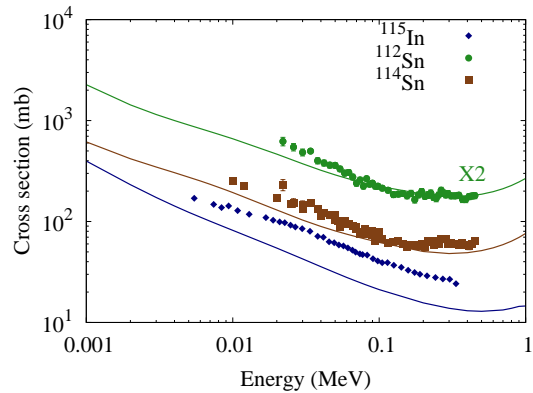}
\caption{Comparison of $(n,\gamma)$ cross sections of the present calculation
with experimental measurements for $^{115}$In and $^{112,114}$Sn. Solid lines indicate  theoretical results.  For  convenience of viewing,  cross section values of $^{112}$Sn are multiplied by a factor of 2.
\label{snngxs1}}
\end{figure}
\begin{figure}
\includegraphics[scale=0.600]{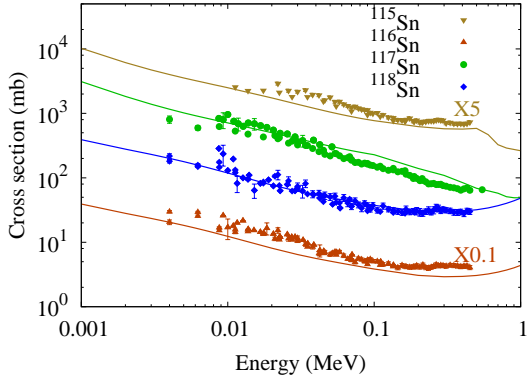}
\caption{   Comparison of $(n,\gamma)$ cross sections of the present calculation
with experimental measurements for $^{115,116,117,118}$Sn. Solid lines indicate the theoretical results.  For  convenience of viewing, cross section values of  $^{115}$Sn and $^{116}$Sn are multiplied by factor of 5 and 0.1, respectively.
\label{snngxs2}}
\end{figure}
\begin{figure}
\includegraphics[scale=0.600]{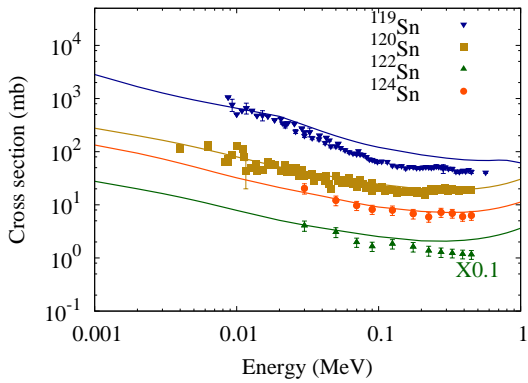}
\caption{   Comparison of $(n,\gamma)$ cross sections of the present calculation
with experimental measurements for $^{119,120,122,124}$Sn. Solid lines indicate the theoretical results. For  convenience of viewing,  cross section values of $^{122}$Sn are multiplied by a factor of 0.1.
\label{snngxs3}}
\end{figure}

\section{Results}
\begin{figure}
\center
\includegraphics[scale=0.600]{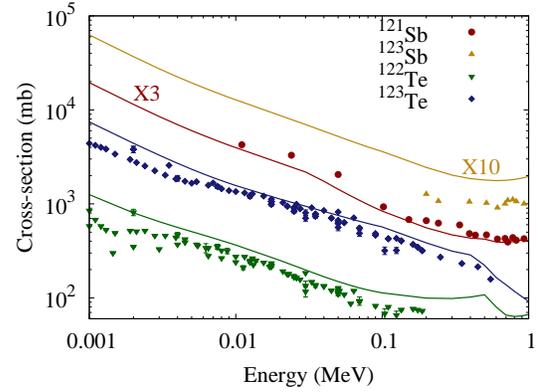}
\caption{   Comparison of $(n,\gamma)$ cross sections of the present calculation
with experimental measurements for $^{121,123}$Sb $^{122,123}$Te. Solid lines indicate the theoretical results. For  convenience of viewing, cross section values of $^{121}$Sb and $^{123}$Sb are multiplied by  factors of 3 and 10, respectively.
\label{tengxs1}}
\end{figure}

\begin{figure}[htp]
\center
\includegraphics[scale=0.600]{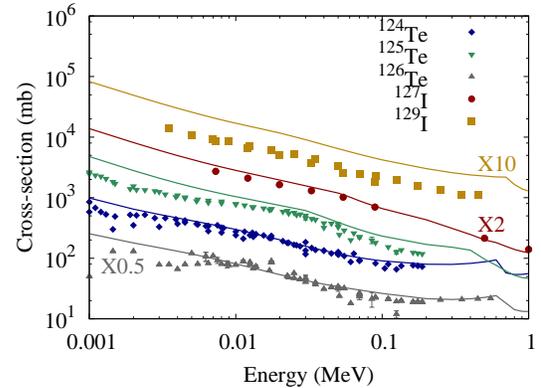}
\caption{Comparison of $(n,\gamma)$ cross sections of the present calculation
with experimental measurements for $^{124-126}$Te. Solid lines indicate the theoretical results. For the convenience of viewing, cross section values of $^{127}$I, $^{129}$I, and $^{126}$Te are multiplied by  factors of 2, 10, and 0.5, respectively.
\label{tengxs2}}
\end{figure}

\begin{figure}
\center
\includegraphics[scale=0.600]{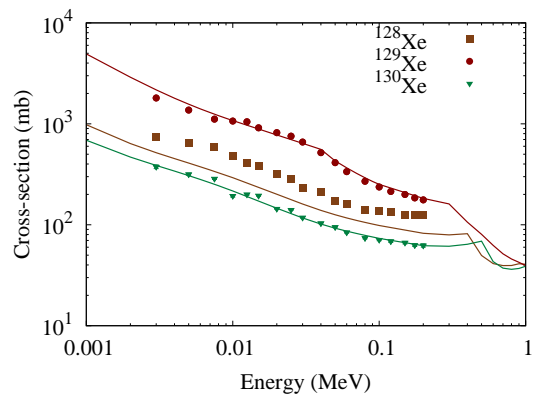}
\caption{   Comparison of $(n,\gamma)$ cross sections of the present calculation
with experimental measurements for $^{128-130}$Xe. Solid lines indicate the theoretical results.
\label{xengxs}}
\end{figure}

Fig.\ref{s-path} shows the s-process nucleosynthesis chain near the $Z=50$ shell closure. 
 We have plotted theoretical $(n,\gamma)$ cross sections with  experimental measurements for a number of $(n,\gamma)$ reactions in the range of interest from Fig. \ref{snngxs1} to Fig. \ref{xengxs}. Experimental values on various targets are available in the website of National Nuclear Data Center \cite{nndc}. In general, the  recent measurements are taken for comparison. In some cases, two or more different data sets are plotted to accommodate a sufficient number of data over the entire energy interval of interest.
 
The theoretical $(n,\gamma)$ cross sections on $^{115}$In and several isotopes of tin
 are plotted with experimental data in Figs. \ref{snngxs1}, \ref{snngxs2}, and \ref{snngxs3}. Experimental data for $^{115}$In is from the measurement of  Kononov \etal \cite{in115_1xs}.
The proton magic element tin  has the greatest number of stable isotopes.  The isotopes $^{112,114}$Sn are very rare and produced only in p-process.
 Abundances are most precisely measured with the isotopes of a single element and hence a large number of stable isotopes of tin provides an opportunity to test the accuracy of theoretical models. 
  The isotope $^{116}$Sn is shielded against r-process $\beta$ decay flow by the stable  isobar $^{116}$Cd. 
Hence, it is least affected by any nearby branchings and  experiences
 complete s-process flow. Hence, it can be used to normalize the entire $\sigma N$ abundance curve. 

 Isotopes $^{114,115}$Sn have very low enrichment. 
 Timokhov \etal \cite{sn112_1sn114_2sn115_2sn122av1sn124av1xs} measured the capture cross sections on stable isotopes of tin including  $^{112,114-120,122,124}$Sn. 
 We have taken the data for $^{118,119}$Sn from Macklin \etal \cite{sn117sn118sn119sn120_1xs}. We have also plotted the data of Nishiyama \etal \cite{sn116_2xs} for $^{116-119}$Sn and Koehler \etal\cite{sn116_1xs} for $^{116}$Sn. These older measurements do not provide data for the low energy region below 20 keV.
 Later, Wisshak \etal \cite{sn114_3sn115_3sn115_1xs} measured neutron capture cross sections  on $^{114-118,120}$Sn from 3 to 225 keV using gold as standard. They reported an uncertainty  $\sim$ 1 \%  which is better than the previous measurements.
 We have also plotted their data with our calculations. As can be seen from Figs. \ref{snngxs1}, \ref{snngxs2}, and \ref {snngxs3} that the data agree fairly with our theoretical values.

Antimony has two stable isotopes, $^{121}$Sb and $^{123}$Sb. This element bypasses the nucleosynthesis flow from tin to tellurium.
 Tolstikov \etal \cite{sb121sb123xs} measured the capture cross sections on $^{121,123}$Sb isotopes via activation technique in the energy range from 
0.3-2.7 MeV. They also compared their results with the statistical model calculations using an optical model potential. We have plotted their measured values with our theoretical calculations in Fig. \ref{tengxs1}.
 
Tellurium, the close neighbor of the element tin, is unique as it has three s-only nuclei among its eight naturally occurring isotopes and one of them has an odd mass number.   The three elements $^{122,123,124}$Te are shielded from r-process by the isotopes of tin and antimony.  The three neighboring s-only isotopes can be used to check the validity of local approximation, i.e., $\sigma_{A}N_{A}$= constant, in classical s-process scenario where flow equilibrium is believed to be achieved.
 The natural abundances are quite low due to the small cross sections near the shell closure. Hence, high sensitivity  in the experimental procedure is demanded. 
Similar to $^{116}$Sn, the isotope $^{124}$Te is also subject to full s-process flow and can become useful calibration point to constrain the main s-process.
 Xia \etal \cite{te123xs1} measured the neutron capture cross sections for $^{122,123,124}$Te from 1 to 60 keV energy range using a setup of three Moxon-Rae detectors and time of flight (TOF) technique.
 The systematic uncertainty was $\sim$ 5\% whereas the
 statistical uncertainty was less than 2\%. Wisshak \etal \cite{te123_2te124_1te125_1te126_1xs} measured the
 neutron capture cross sections on $^{120,122-126}$Te in the energy range from 10 to 200 keV. 

 Previously, Macklin and Gibbons \cite{te122te123_3te124_3te125_3xs} used total
 energy technique to determine cross sections on the same
 isotopes of tellurium from 30 to 220 keV.
We have plotted these experimental data with our calculated results in Figs. 
\ref {tengxs1} and \ref {tengxs2}. 
The $(n,\gamma)$ cross sections for $^{127,129}$I are plotted in Fig. \ref{tengxs2}. 
  The isotope $^{129}$I is a long-lived fission product with $\beta$ decay half-life of about 15.7 million years  and is useful in nuclear transmutation technology. Its formation in s-process is blocked by the instability of $^{128}$I. It can be formed in r-process via the $\beta$ decay of $^{129}$Te. 
Noguere \etal \cite{i127_2i129_2xs} have recently measured the $(n,\gamma)$ cross sections on $^{127,129}$I.  We have also plotted the data of  Voignier \etal \cite{i127_1xs} for $^{127}$I and Macklin \cite{i129_1xs} for $^{129}$I.

The element xenon is of particular interest as it has nine stable isotopes produced in different processes under heavy element nucleosynthesis. 
 It has two s-only isotopes $^{128,130}$Xe 
Unlike other elements, solar xenon
 abundance  can not be determined from the analysis of spectral meteorites or primitive
 meteorites. Thus, it has to be determined  from the systematic
 study of $\sigma$N statistics. 

 Fig.~\ref{xengxs} shows 
the calculated $(n,\gamma)$ cross sections for $^{128,129,130}$Xe. 
Experimental data are taken from Reifarth \etal \cite{xe128xe129xe130xs}. 
They have measured the $(n,\gamma)$ cross sections using gold as standard in 
the neutron energy range from 3 to 225 keV using TOF method with an uncertainty
 of 2\%. The ratios of xenon to gold experimental cross sections were converted into absolute xenon cross sections by using gold data of R. L. Macklin (private communication) followed by a normalization by a factor of 0.989  with the absolute value of Ratynski and K{\"a}ppeler \cite{ratynski_gold_ref}.
\setlength{\tabcolsep}{10pt}
\renewcommand{\arraystretch}{1.2}
\begin{table*}[htb]
\center
\caption{Maxwellian averaged cross sections at $kT=30$ keV for nuclei near the 
$Z=50$ shell closure. Experimental values are from Ref. \cite{kadonis1}.
 For unstable and radioactive nuclei, experimental data are not available. See text for other 
experimental values. 
\label{macs30kev}}
\begin{tabular}{crrrcrrr}\hline
 &\multicolumn{3}{c}{MACS (mb)}&
 &\multicolumn{3}{c}{MACS (mb)}\\\cline{2-4}\cline{6-8}
Nucleus&Present & Exp. & MOST & Nucleus&Present & Exp. & MOST\\
\hline
$_{49}^{113}$In&603&787$\pm$70&314&
$_{49}^{115}$In&433&706$\pm$70&298\\
$_{50}^{112}$Sn&188.7 &210$\pm$12  &153  &
$_{50}^{114}$Sn&101.5& 134.4$\pm$1.8 &  72.9\\
$_{50}^{115}$Sn&284& 342.4$\pm$8.7 &  245&
$_{50}^{116}$Sn&69& 91.6$\pm$0.6 &  45.6\\
$_{50}^{117}$Sn&415& 318.8$\pm$4.8 &  299&
$_{50}^{118}$Sn&66.3& 62.1$\pm$0.6 &  48.2\\
$_{50}^{119}$Sn&306& 180$\pm$10 &  240&
$_{50}^{120}$Sn&44.3&36.2$\pm$0.3 &  30.2\\
$_{50}^{121}$Sn&217& &  341&
$_{50}^{122}$Sn&44.5&21.9$\pm$1.5 &  32.9\\
$_{50}^{124}$Sn&17.6&12.0$\pm$1.8 & 13.6\\
$_{51}^{121}$Sb&678& 532$\pm$16 &  417&
$_{51}^{122}$Sb&1229& &  772\\
$_{51}^{123}$Sb&709 &303$\pm$9 &  422\\
$_{52}^{122}$Te&204 &295$\pm$3 &  165&
$_{52}^{123}$Te&981 &832$\pm$89 & 548\\
$_{52}^{124}$Te&164 &155$\pm$2 &  105&
$_{52}^{125}$Te&552 &431$\pm$4 &  382\\
$_{52}^{126}$Te&91.3 &81.3$\pm$1.4 & 73.7&
$_{52}^{127}$Te&851&&\\
$_{52}^{128}$Te&101&44.4$\pm$1.3 &74.3\\
$_{53}^{127}$I&766&635$\pm$30 & 470&
$_{53}^{128}$I&1407.2& & \\
$_{53}^{129}$I&901&441$\pm$22 &497\\
$_{54}^{124}$Xe&692 &644$\pm$83&500&
$_{54}^{126}$Xe&471&359$\pm$51&333\\
$_{54}^{128}$Xe&169&262.5$\pm$3.7&105&
$_{54}^{129}$Xe&588&617$\pm$12&307\\
$_{54}^{130}$Xe&125&132.0$\pm$2.1&89.8&
$_{54}^{131}$Xe&618 &&319\\
$_{54}^{132}$Xe&60.1&64.6$\pm$5.3&67.4&
$_{54}^{133}$Xe&464&&235\\
\hline
\end{tabular}
\end{table*}
We have also calculated the Maxwellian averaged cross sections and compared them with the recommended values by Bao \etal \cite{bao}  and theoretical MOST2005 \cite{most2005} predictions in Table~\ref{macs30kev}. These values are available  in KADoNiS data base \cite{kadonis1}.

\setlength{\tabcolsep}{12pt}
\renewcommand{\arraystretch}{1.25}
\begin{table*}
\center
\caption{Maxwellian-averaged cross sections over a range of thermal energies for several branch-point nuclei  in the s-path near the Sn shell closure and for stable $^{131}$Xe.
\label{macs_range}}
\begin{tabular}{cccccc}\hline
$KT$ (MeV)&\multicolumn{5}{c}{MACS (mb)}\\\cline{2-6}
&$^{121}$Sn$(n,\gamma)$ & $^{122}$Sb$(n,\gamma)$ &$^{127}$Te$(n,\gamma)$&$^{131}$Xe$(n,\gamma)$&$^{133}$Xe$(n,\gamma)$\\\hline
0.005&683&3601&2400&1778&1348\\
0.010&437&2346&1565&1159& 873\\
0.015&338&1855&1241& 915& 686\\
0.020&281&1572&1060& 777& 582\\
0.025&244&1377& 939& 685& 514\\
0.030&217&1229& 851& 618& 464\\
0.040&179&1012& 727& 527& 398\\
0.050&154& 859& 643& 466& 353\\
0.060&137& 744& 582& 421& 421\\
0.080&113& 583& 499& 359& 277\\
0.100& 98& 476& 443& 316& 248\\
\hline
\end{tabular}
\end{table*}
\setlength{\tabcolsep}{10pt}
\renewcommand{\arraystretch}{1.25}

\begin{table*}
\center
\caption{Astrophysical reaction rates  $N_{A}<\sigma v>$ (cm$^{3}$ mol$^{-1}$ sec$^{-1}$) over a range of stellar temperature from the present calculations for the targets 
$^{112,114,115}$Sn and $^{131}$Xe. The rates are in the order of $10^{7}$. For the sake of comparison, we have also listed the theoretical rates from BRUSLIB data base \cite{bruslib}.
\label{rates}}
\begin{tabular}{ccc|cc|cc|cc}\hline
$T_{9}$(GK) & \multicolumn{8}{c}{$N_{A}<\sigma v>$ (cm$^{3}$ mol$^{-1}$ sec$^{-1}$)}\\\hline
&\multicolumn{2}{c}{$^{112}$Sn}&\multicolumn{2}{c}{$^{114}$Sn}
&\multicolumn{2}{c}{$^{115}$Sn}&\multicolumn{2}{c}{$^{131}$Xe}\\\cline{2-3}\cline{4-5}\cline{6-7}\cline{8-9}
& Pres. & Ref.\cite{bruslib}& Pres.& Ref.\cite{bruslib}& Pres. & Ref.\cite{bruslib}
& Pres. & Ref.\cite{bruslib}\\\hline
0.1&2.754&3.308&1.594&1.405&4.273&3.239&9.831&4.473\\
0.2&2.746&3.287&1.562&1.309&4.132&3.128&9.250&4.291\\
0.3&2.785&3.250&1.571&1.263&4.110&3.045&8.960&4.186\\
0.4&2.829&3.222&1.588&1.241&4.118&2.983&8.874&4.153\\
0.5&2.877&3.206&1.608&1.232&4.138&2.939&9.003&4.206\\
0.6&2.928&3.205&1.631&1.234&4.169&2.913&9.274&4.316\\
0.7&2.982&3.219&1.657&1.244&4.210&2.901&9.598&4.445\\
0.8&3.041&3.247&1.686&1.259&4.260&2.901&9.912&4.567\\
0.9&3.104&3.286&1.718&1.279&4.319&2.913&10.827&4.671\\
1.0&3.171&3.335&1.752&1.303&4.389&2.934&10.398&4.751\\
2.0&4.043&4.189&2.228&1.662&5.631&3.551&10.379&4.477\\
3.0&5.389&5.597&3.018&2.203&6.652&3.640&7.024&1.912\\
4.0&6.954&6.647&3.878&2.367&5.304&1.849&2.446&0.522\\
5.0&4.971&4.296&3.647&1.359&2.287&0.707&0.645&0.189\\\hline
\end{tabular}
\end{table*}

For a number of selected isotopes, we have presented the Maxwellian-averaged cross sections and reaction rates over a range of astrophysical   energies and  temperatures in Tables  \ref{macs_range} and \ref{rates}.  The neutron capture cross sections of isotopes acting as important branch points plays crucial role in the production of a few s-oly isotopes. Moreover, these unstable branch points are barely accessible to experimental measurements. The production of three neighboring s-only isotopes $^{122,123,124}$Te is governed by the branches at $^{121}$Sn and $^{122}$Sb. The branch at $^{127}$Te has contribution towards the production of s-only isotopes $^{128,130}$Xe, the MACSs of which are well-determined with uncertainty less than 2\%. The branch at $^{133}$Xe  has a marginal  neutron flow towards neutron-rich isotope $^{134}$Xe owing to its short $\beta$-decay life time ($t_{1/2}=5.2$ days).  The two important s-only isotopes $^{134,136}$Ba  are affected by branch-points those are strongly sensitive to stellar temperature, electron density and neutron density. The MACS values for the branch-point nucleus 
$^{133}$Xe is needed to correctly predict the small but non-negligible s-predictions for these two s-only nuclei. These four unstable short-lived branch-point isotopes do not have  experimental data. The isotope $^{127}$Te even has no theoretical estimation made so far for MACS. Hence, we have presented MACS values for them from 5 to 100 keV.  The Experiment has yet not been performed on the stable isotope $^{131}$Xe, possibly due to low enrichment  in sample. Hence, we have presented our calculated values for both cross sections and reaction rates over a range of thermal energies and stellar temperatures for this nucleus.

The astrophysical origin of some rare isotopes, e.g., $^{112,114,115}$Sn in this region of interest has been a long-standing problem. These nuclei are  bypassed by the main flow  of s-process and  post freeze-out $\beta$ decay flow of r-process and hence,  are classified as p-nuclei. Their abundances are small and it is difficult to predict exact s-process contributions towards their abundances. The empirical abundance scaling laws have been proposed between the p- and s-only nuclei with the same atomic number in solar abundances from theoretical model calculations \cite{hayakawa}. These scaling laws suggest that they are basically produced in photo-disintegration reactions. The rates of such reactions can be obtained from the forward $(n,\gamma)$ rates by the principle of detailed balance. For this reason, we have presented the astrophysical radiative neutron capture  rates from our  calculations for the rare p-only isotopes of tin in Table \ref{rates}. The theoretical rates from the BRUSLIB data library \cite{bruslib} are also listed alongside to show the difference. The rates in the BRUSLIB data base were evaluated using the reaction code TALYS. The inputs for statistical model calculations are different from our study. Various global and local microscopic models were used for nuclear structure properties, deformation, optical potential, level densities, strength functions, etc., whenever the experimental information were not available. The optical potential of Ref. \cite{bruslib_optmod} was used. Nuclear masses and deformations were from HFB mass model  based on  extended Skyrme force with a four parameter delta function pairing force \cite{bruslib_hfb}. The E1 $\gamma$-ray strength functions were taken from HFB+QRPA calculations of Goriely \cite{bruslib_e1}. Nuclear level densities were taken from the same reference as  in our present case of study. It can be seen that the rates that our calculations yield are quite different from those of BRUSLIB data base. Especially for stable $^{131}$Xe, our predicted values are larger by  more than 2 times at comparatively lower temperature and by more than 3 times at temperatures $>$ 1 GK.

\section{Discussion and Summary}
From the figures and tables it can be observed that our microscopic model reproduces the experimental cross section values for nuclei studied in the present case reasonably, except for a very few cases. Our theory yields almost twice the values for $^{122}$Sn, $^{123}$Sb, $^{129}$I, and $^{128}$Te while that of $^{115}$In is a factor of half less than the experimental one at 30 keV. 
 Nevertheless, agreements between theory and experiment for MACS values are within the satisfactory limits with the difference ranging in between $\sim$ 5-25\%. For example, experimentally the cross sections of $^{124}$Te has been known with accuracy of 1\%. We present a value that differs from this by 5\%. This definitely depicts the predictive power of our theoretical model.  The exceptions are $^{115}$In, $^{128}$Te, $^{129}$I, for which our model either largely overestimates or underestimates the measurements. It is notable as well as interesting that for $^{122}$Sb and $^{131,133}$Xe the predicted MACS values at 30 keV are very different from the pre-existing  MOST calculations. 
It is to be noted that most of the nuclei studied in the present work near the neighborhood of the $Z=50$ shell-closure and participating in the main s-component reside close to the stability valley. Hence, deformation is very small and the spherical optical
 model described in terms of nuclear radius $R=r_{0}A^{1/3}$ can be reliably applied. However in case of small quadrupole or hexadecapole deformations arising due to static or dynamic collective rotational or vibrational excitations should be more accurately treated in Distorted-wave Born Approximation (SWBA) calculation or  by coupled channel calculations in case  of strong deformations.

Application of statistical model requires a large number of overlapping resonances at the compound nucleus formation energy so that the average of the transmission coefficients, those do not show resonant behavior, can be obtained. This, in turn, requires a large number of levels per energy range in the compound nucleus which can act as doorway states to its formation. The individual resonance widths then  can be replaced by an average one. However, the nuclei near the shell closures, in general,  have low level densities and it is difficult to apply the statistical model in these regions. However, a comparatively broader s-wave neutron level spacings allow one to apply statistical model at low level densities even near the shell-closures. Statistical model predicts somewhat overestimated values of cross sections for  low  level densities in compound nuclei \cite{ldens_low}. A lower limit to the number of levels per energy window that is enough to replace the sum by an integral over the HF cross section is set to be about 10 in the worst case, by the numerical calculation in Ref. \cite{rau1}. Rauscher \etal \cite{rau1} have also derived and presented a lower temperature limit  above which  the statistical model reaction rate calculations for various neutron, proton, and $\alpha$-induced reactions are valid.
At low energies of astrophysical regimes, the cross sections can have direct capture component as well as contribution from individual or narrow single resonances those have to be described by Breight-Wigner terms. In extreme cases, interference terms may also appear. 

Within stars, nuclei exist both in the ground state and thermally populated excited states in statistical equilibrium. Hence, cross sections for nuclei existing only in the ground state gives merely an incomplete picture. To get the accurate stellar cross sections, these values have to be complemented with  stellar enhancement factors (SEFs).
\begin{equation}
SEF=\frac{< \sigma>^{*}}{<\sigma>^{gr}}
\end{equation}
Here,  $<\sigma>^{*}$ denotes the MACS values averaged over thermally populated levels for nuclei within stars  and $<\sigma>^{gr}$ denotes the MACS values for nuclei in the ground state.
 However, ground state cross sections can easily be compared to the laboratory measurements. Our cross sections do not take these SEFs into account. However, at low energies and temperatures relevant to neutron capture reactions for s-process, especially for the main component in low mass AGB stars, these SEFs are below 8\%. Hence, we do not expect significant modifications or changes in our results after the inclusion of these factors. Moreover, due to the fact that the ground state cross section constitutes only a minor fraction of MACS, uncertainties in theoretical MACS values may be in some cases largely underestimated. 

It is to be noted that we have provided the cross sections and reaction rates from a complete theoretical view point. We have not used any experimental information for the nuclear inputs (except for nuclear binding energies). Even, the nuclear density distributions for folding the effective interaction has been taken from theoretical relativistic-mean-field model.  The important inputs, such as  level densities and E1-$\gamma$-ray strength functions are taken  from  current microscopic models which are considered to be more reliable and hence can better predict the observables away from the experimentally accessible region in the nuclear landscape than other local or global phenomenological models. It has been seen that the largest uncertainty in theoretical calculations results from an inappropriate description of nuclear level density. Obviously, if one uses experimental information, a better accuracy could be achieved.  However, a complete theoretical formalism makes our model acceptable to predict the values for  nuclei away from stability valley in the nuclear landscape for which experimental information is scarce or still do not exist. Moreover, we have not locally tuned the parameters for each individual reaction in doing which the agreement between theory and experiments would certainly  become  better. However, as our aim is to apply this model to predict those values for which cross sections or rates are still unknown, a single  parameterized  approach over entire energy region and mass range is nevertheless more convenient. 
Thus,  it can be concluded that provided our statistical model has some limitations, it is moderate to predict the unknown values reasonably within a certain range of astrophysical energies and temperatures.

In summary, we have studied the $(n,\gamma)$ cross sections  for nuclei  of astrophysical importance theoretically and compared them with the available experimental measurements.
The DDM3Y NN interaction, folded with target radial matter densities, obtained from relativistic mean field approach has been used to construct  the microscopic  optical 
model potential. Finally, we have presented Maxwellian-averaged cross sections and astrophysical reaction rates from the prediction of our theoretical model.

\section{Acknowledgement}
Authors acknowledge  the University Grants Commission (Junior Research Fellowship and Departmental Research Scheme) and the Alexander Von Humboldt Foundation for providing financial assistance.

\end{document}